# Direct Foreign Investment in Kurdistan Region of Middle-East: Non-Oil Sector Analysis

**Angus Okechukwu Unegbu[1*] and Augustine Okanlawon[1]**

[1]*Department of Business and Management Sciences, University of Kurdistan Hewler, Iraq.*

***Authors' contributions***

*This work was carried out in collaboration between all authors. Author AOU designed the study, wrote the protocol, the first draft of the manuscript, carried out the literature searches, analyzes the study while author AO obtained the research data and did the preliminary editorial works. All authors read and approved the final manuscript.*

***Article Information***

DOI: 10.9734/BJEMT/2015/14001
Editor(s):
(1) Chen Zhan-Ming, School of Economics, Renmin University of China, Beijing, China.
Reviewers:
(1) Abdulsalam Mas'ud, Department Ofaccounting, Hussaini Adamu Federal Polytechnic, Kazaure –Nigeria/ School Of Accounting, Universiti Utara Malaysia, Malaysia.
(2) George O. Tasie, School of Business and Entrepreneurship, American University of Nigeria, Yola, Adamawa State, Nigeria.
Complete Peer review History: http://www.sciencedomain.org/review-history.php?iid=811&id=20&aid=7126

***Original Research Article***

***Received 14th September 2014***
***Accepted 4th October 2014***
***Published 8th December 2014***

## ABSTRACT

Kurdistan Region is a tourist hub. This research analyzes other Non-Oil Sectors that have huge attractions of Foreign Direct Investments into the Kurdistan Region from 2005 to 2013. Comparative analysis was carried out between Iraq and the Region, and among influential Sectors of the Economy. T-test and ANOVA are statistical tools employed in testing the research hypotheses. The research identify that there exist significant Foreign Direct Investment inflows across the governorates in the region and among influential sectors of the Economy. The research also highlighted areas of high level of investment needs, sectors that have been crowded out and business opportunities in the region that requires huge Foreign Direct Investments. It is recommended that the Regional Kurdistan Government should embark on fiscal Cashless policies in order to stimulate further spill-off effects of attracting enormous Non-Oil Sectors of Foreign Direct Investments into the region.

---

**Corresponding author: E-mail: unegbu4@yahoo.com;*

## 1. INTRODUCTION

Kurdistan is a region in a decentralized Iraqi federation with asymmetric options [1] asserts. This assertion has a legal (Iraqi 2005 Constitution) backing that affirms Kurdistan Region as such in the Iraqi federation. It is good to note that Iraq is not constitutional based on a political system of ethnic territorial units with overlapping political administrative entities, but if the Kurdistan push succeeds; it may be argued that the form of Iraqi federalism will be constructed around 'Ethnic Federalism'. The quasi-autonomous status empowered on the region legally and politically has spread to attracting massive direct foreign investment. Within the last ten years, more than 40b dollar Direct Foreign Investment (DFI) came to the region. DFI (Also refer as Foreign Direct Investment - FDI) by multinational corporations, Western Countries and other neighboring Countries are huge in the Oil Sector of the economy. Huge and massive direct foreign investment in the oil sector created awesome positive externalities in attracting foreign technology, human capital and expansion of Direct Foreign Investment into non-oil sectors of the economy. In 2012 alone according to [2] 128 licenses were granted for Foreign ventures into the Oil sector with a potential value of over $6.3b in Iraq, out of which 67% went to Kurdistan Regional Government. This and other huge Direct Foreign Investments in the Oil Sector will naturally create big ripple effect in attracting massive Direct Foreign Investment into the Non-Oil Sectors of Kurdistan Region.

The cut-off period from 2005 to 2013 figures are essentials as 2005 marks the recognition of Kurdistan as Regional Government with quasi-autonomous status. Table 1 below confirms that since 2005 there is not only massive Foreign Direct Investment in Iraq but there exist huge increasing Foreign Direct Investment from 2005 to 2013.

It is recorded by [3] that Foreign Direct Investment in Iraq for the years covering 2005 to 2013 are as seen in Table 1.

The subject of Foreign Direct Investment from 2005 to 2013 in Iraq has been addressed in several studies [4,5,2,6,3,7,8].

It is pertinent to observe that out of these many studies, there is a dearth in research on Direct Foreign Investment in the Kurdistan Region – host to vast oil and gas reservoirs with high openness to foreign investments.

### 1.1 Established Research on Causality Relationship between FDI and Oil Exploration

It is reasonable to assert that, there is causality relationship between Foreign Direct Investment (FDI) existing on large economic activities of Oil & Gas to growth on Non-Oil Sectors of Economies. In the words of [9] causes of Foreign Direct Investment inflow to domestic economy among other factors include neighborhood externalities and the presence of natural resources such as Oil and Gas reservoirs. In another aspect [10] establish causality relationship between FDI to Non-oil economic activity growth in an Oil-dominated economy. In their research, using econometric analysis, they were able to establish that in the long-run, FDI affects economic growth on non-oil sectors positively. DFI or FDI economic and technological advantages are well documented by [11,12,13]. It is also the assertion of [11] that FDI has proved to stimulate growth and development of countries. Kurdistan Region is not an exception. FDI is also a source of acquiring valuable technological know-how, apart from fostering linkages with other firms, which can jumpstart economic activities. This means that a spillover effect from FDI on Oil drilling and other economic activities are readily common to Non-oil Sectors of the Economy. To support this line of thought [14] opine that spillovers are traditionally expected to accrue to the industry by mere entry of multinational entities. Other studies on the determinants of Foreign Direct Investments exist according to [15,4,10,16]. Expectedly, the bulk of these determinants of FDI in Oil find Countries is Oil and Oil related component. However, worthy of note in all the determinants is that [15] build in a very vital variable in their indicators of FDI determinants – Openness of Trade, though there exist controversy as assert by [17,18,15,14,19] as to what extent openness to trade has a positive correlation between Gross Domestic Product growths with the level of FDI to the domestic economy. Compared with and between Governments' policies across regions/countries intra and inter Kurdistan Regional Government's policies, Kurdistan should be adored as exhibiting openness to Foreign Investment. According to statistics released [20], the region has attracted some $16.2 billion (19 trillion IQD) in foreign investment over the past five years.

### 1.2 Non-Oil Sectors in Kurdistan Region

Investment in Kurdistan Region is guided by Kurdistan Region Investment Law (Law 4 of 2006). It is this law that empowers the Kurdistan Board of Investment. Kurdistan Board of Investment as documented by [18] outlines the relevant Sectors as such as; Housing, Tourism, Agriculture, Health, Education, Banking, Communication & Transport, Trade & Industry and Arts & Sports.

The Housing Sector development according to [18] has fuelled a huge rise in housing demands thus contributing to economic growth. The need for more housing in the region is also an offshoot of increasing number of Companies located in the region. It is the view of [18] that as at date, 78 housing projects amounting to more than $5.8 billion had been completed within the region. The locations, capital investments and Project Centers as shown in Table 2.

Tourism Sector in Kurdistan is a boom resulting from stability, accommodation policies and openness to modernizations. Also contributing to tourism boom in the region are historical sites and natural sceneries Pank Tourism Complex in Rawaduz. The scenery is a beauty and friendly strange to Middle East hostility. It is asserted by (21) that there were 2, 216,993 tourists in the region in 2012 and over 259 hotels.

Agricultural Sector although disadvantaged, is making inroads because Kurdistan region has arable and fertile land, conducive climate, strong human resources and plentiful water resources according to Kurdistan Board of Investment. A Kurdistan Regional area of agricultural importance is shown in Table 3.

Health Sector in Kurdistan region is faced with challenges but according to [18] over one hundred and ninety-five million dollars have been invested on seventeen projects in the region, the purpose of which is to enhance and overcome some of the challenges.

Education Sector in the region is massively being restructured to enhance Private Sector involvement and this is yielding huge entry. There are a good number of Universities in the region, although there exist few English-taught Universities in the Region. Data obtained from Kurdistan Board of Investment shows that more than three hundred and seventy five million dollars have been invested into the sector as at date.

Banking and Insurance Sector attracted and is still attracting huge capital investment. In the region there are two State owned Central Banks, fourteen State owned Banks and thirty privately owned Banks. As at date, there is over $2.3billion investment in the Banking and Insurance Sector. Although, this witnessed large growth in banking does not yet has same correlation to Insurance in the region.

**Table 1. Direct foreign investment in Iraq from 2005 to 2013**

| FDI (USD Million) | 2005 | 2006 | 2007 | 2008 | 2009 | 2010 | 2011 | 2012 | 2013 |
|---|---|---|---|---|---|---|---|---|---|
| FDI Inflow | 515 | 383 | 972 | 1,856 | 1,598 | 1,396 | 2,082 | 2,549 | N/A |

*Source: International Trade centre*

**Table 2. Housing sector projects and locations in Kurdistan**

| Project center | Location | | |
|---|---|---|---|
| | Erbil | Duhok | Sulaymaniah |
| American village | $80,000,000 | | |
| Avro city | | $500,000,000 | |
| Dream city | $300,000,000 | | |
| Ganjancity | $312,000,000 | | |
| German city | | | $62,000,000 |
| Naz city | N/A | | |
| PAK city | | | N/A |
| English village | N/A | | |

*Source: Kurdistan Board of Investment*

**Table 3. Kurdistan Regional areas of agricultural importance**

| Regional area/zone | Agricultural importance |
|---|---|
| Sharezoor valley | Vegetables, diary and feed production. |
| Kalar | All round agricultural produce (Multiple projects). |
| Rania plains | All round agricultural produce (Multiple projects). |
| Erbil plains | Grain production, feed and vegetable production, poultry and cattle diary. |
| Hareer plains | Vegetables and oil production. |
| Aqra plains | Rice, grain and Vegetables, poultry and dairy farms. |
| Zakho plains | Grain and livestock |

*Source: Kurdistan Board of Investment*

Communication and Transport Sector in Kurdistan is having a field day. There are two International Airports – Erbil International Airport and Sulaymaniah International Airport in the Region. There is considerable advancement in telecommunications as the Region is a host to two mobile phone providers.

Trade and Industrial Sector in the region as at date is within the range of $5billion annually as posits by [18]. The region is consumer import orientated.

The Arts and Sports Sector in Kurdistan Region despite historical existence of many edifices, begs for real capital and human investment.

The purpose of this research is to conduct analytical study of Direct Foreign Investment in Kurdistan Region with a specific objective of carrying out Non-oil Sector analysis from 2005 to 2013. The relevance of this research, among others is that it contributes to literature on opportunities of investment in Kurdistan mainly to non-oil sectors of the economy, showing sectorial areas that have been crowded out. It will also analyze influence of Oil on Direct Foreign Investment to non-oil sectors using quantitative facts, the outcome contributes to professional knowledge and sets pace for further studies in this sphere. The research will also contribute in laying a foundation for database in a region where data collection and dissemination is material for further economic growth is lacking.

In order to drive home the specific objectives, the following null-hypotheses are hereby formulated;

1. None of the comparative analyses of; Sectors verses Amount of FDI, Governorates verses Amount of FDI, Sectors verses Number of FDI and Governorates verses Number of FDI in Kurdistan Regional from 2005 to 2013 is significant.
2. FDI in Kurdistan Region is not significantly different compared with that of Iraq between the periods of 2005 to 2013.
3. The level of Foreign Direct Investments between Non-Oil Sectors in Kurdistan Region from 2005 to 2013 is not significantly different.

This study is structured to introduce the title and justifying its significance. The introduction is garnished with a brief review of existing relevant literature, after which the research method adopted is highlighted. Data presentations, Analysis, Results and Discussions and Findings will follow. Conclusions and Recommendations end the research design.

## 2. RESEARCH METHODS AND DATA

The data collection method adopted is statistical exploration of existing data from established sources aimed at establishing a data base and carrying out relevant analysis of Non-Oil Sector Foreign Direct Investment in the Kurdistan Region from 2005 to 2013. Comparative analytical studies between Iraq and Kurdistan Regional Government is quantitatively analyzed using available data to test stated hypotheses.

The data for this research were collected from numerous publications of Kurdistan Region Board of Investment, Studies and Information Department, though inherent human errors should be attributed to the Researchers.

A compendium of these publications Kurdistan Region Board of Investment were professional organized and analyses were made by sectorial classifications, thus highlighting areas of high need, moderate need, low need and areas that have been crowded out. Percentages are used to highlight areas of significance. The research hypotheses were tested using SPSS software Paired T-test for hypothesis 1 and 2, while ANOVA was used for hypothesis 3 testing. Decision criteria for acceptance or rejection of the outcomes of these hypotheses are based on 95% confidence levels, thus test statistics of equal or lower than 5% is considered significant.

## 3. DATA PRESENTATIONS, ANALYSIS, RESULTS AND DISCUSSIONS

### 3.1 Test of Hypotheses 1

To test null-hypothesis 1, which states that none of the comparative analyses of Sectors verses Amount of FDI, Governorates verses Amount of FDI, Sectors verses Number of FDI and Governorates verses Number of FDI in Kurdistan Regional from 2005 to 2013 is significant, the data in Table 4 is used. The relevant statistical tool used is Paired Sample T-test using SPSS software. The outcome appears as seen in Table 8.

The outcome of the test shows that the Amount of Foreign Direct Investment in Kurdistan in all the Sectors from 2005 to 2013 are significantly high as the test result showed that $p<0.05$ with a test static of $p = 0.003$. Same conclusion is also reached for amount of Foreign Direct Investment in a three Governorates of Duhok, Suilamaniya and Erbil.

The Number of Investments among the Sectors are also found to be significant with a test static of $p = 0.004$ which is less than 0.05. As number of Foreign Direct Investment among the Governorates, it was also found to be significantly high with a test static of $p =0.0001$.

Joint-Venture investments between Kurdistan National and Foreigners from 2013 to 2005 among the governorates are very high as seen in Table 5, though it is outside the main focus of this research.

In conclusion, the test outcomes show that hypothesis 1 which states that none of the comparative analyses of Sectors verses Amount of FDI, Governorates verses Amount of FDI, Sectors verses Number of FDI and Governorates verses Number of FDI in Kurdistan Regional from 2005 to 2013 is significant is hereby rejected. A comparative analyses show that the level of Foreign Direct Investment among the Sectors, Governorates and Numbers in Kurdistan Regional Government from 2005 to 2013 are significantly high.

Though the result shows a high significant level across the relevant fields, but the highest number of FDI in the first three Sectors are in Industrial, Housing and Tourism respectively. In terms of the amount of FDI, the order of first three Sectors are; Housing, Industry and Tourism.

Table 6 as shown above captures the investing countries and their percentage of investment. The table shows that the neighbors of Kurdistan Regional Governments like Iraq, Iran, Turkey, Lebanon, and Egypt have high percentage of stake. United Arab Emirate has outstanding level of Foreign Direct Investment in Kurdistan. The table shows also that combinations of Foreign Direct Investments by European Countries are high. Also very important to deduce from Table 6, is the level of Joint-Venture Investment in Kurdistan Region from 2005 to 2013 is enormously huge. Joint-Venture Investment in the region amounted to 8.721%, seconded by United Arab Emirate Investment of above 6.12%.

### 3.2 Test of Hypothesis 2

Hypothesis 2 states that FDI in Kurdistan Region is not significantly different compared with that of Iraq between the periods of 2005 to 2013. To test this hypothesis, a Paired T-test of FDI in Iraq against that of Kurdistan in terms of amount of investments and year of investment are carried out, using available comparative data from Tables 1 and 7. The result is as shown in Table 9.

The outcomes of the test shows that Foreign Direct Investment in the Kurdistan Region and Iraq between the years of 2005 to 2013 are significantly different between the two and in amount with test statics of $p = 0.003$ and 0.0001 respectively. With the outcomes of this test, the hypothesis which states that FDI in Kurdistan Region is not significantly different compared with that of Iraq between the periods of 2005 to 2013 is hereby rejected as test results are both less than 0.05.

### 3.3 Test of Hypothesis 3

Hypothesis 3 states that levels of Foreign Direct Investments between Non-Oil Sectors in Kurdistan Region from 2005 to 2013 are not significantly different. This test is a confirmatory of part outcome of hypothesis 1. The hypothesis is concerned with finding out the significant levels between the Sectors, Governorates and Numbers of Foreign Direct Investments in Kurdistan from 2005 to 2013, thus the best statistical tool to use in this regard is the ANOVA. The outcome is shown in Table 10.

The outcome of hypothesis 3 test shows that the differences in Foreign Direct Investments within the Sectors in Kurdistan Region from 2005 to 2013 are not significant as test static of p =0.887.

**Table 4. Licensed authorized DFI in Kurdistan from 2005 to 2013- sectorial and governorates analyses**

| Sector | Governorates | Number | Investment amount in $ | Total number | Total capital investment ($) |
|---|---|---|---|---|---|
| Agriculture | Duhok | 6 | 431,142,972 | 26 | 704,934.181 |
| | Sulaimaniya | 2 | 10,649,500 | | |
| | Erbil | 18 | 263,141,709 | | |
| Art | Duhok | 0 | 0 | 4 | 11,756,498 |
| | Sulaimaniya | 0 | 0 | | |
| | Erbil | 4 | 11,756,498 | | |
| Banks | Duhok | 0 | 0 | 2 | 740,000,000 |
| | Sulaimaniya | 0 | 0 | | |
| | Erbil | 2 | 740,000,000 | | |
| Communication | Duhok | 0 | 0 | 5 | 220,890,942 |
| | Sulaimaniya | 3 | 92,995,942 | | |
| | Erbil | 2 | 127,895,000 | | |
| Education | Duhok | 7 | 28,960,682 | 19 | 719,601,569 |
| | Sulaimaniya | 5 | 461,547,150 | | |
| | Erbil | 7 | 229,093,737 | | |
| Health | Duhok | 7 | 35,722,954 | 41 | 1,023,116,475 |
| | Sulaimaniya | 3 | 106,411,446 | | |
| | Erbil | 31 | 880,982,075 | | |
| Housing | Duhok | 38 | 2,137,367,598 | 164 | 13,614,320,225 |
| | Sulaimaniya | 45 | 2,422,655,178 | | |
| | Erbil | 81 | 9,054,297,449 | | |
| Industry | Duhok | 57 | 1,241,597,177 | 184 | 12,779,506,218 |
| | Sulaimaniya | 51 | 7,432,125,400 | | |
| | Erbil | 76 | 4,105,783,641 | | |
| Service | Duhok | 0 | 0 | 7 | 188,982,715 |
| | Sulaimaniya | 4 | 89,291,555 | | |
| | Erbil | 3 | 99,691,160 | | |
| Sports | Duhok | 8 | 61,468,932 | 20 | 92,398,998 |
| | Sulaimaniya | 11 | 19,930,066 | | |
| | Erbil | 1 | 11,000,000 | | |
| Tourism | Duhok | 40 | 719,782,751 | 127 | 6,578,363,224 |
| | Sulaimaniya | 16 | 1,391,400,820 | | |
| | Erbil | 71 | 4,467,179,653 | | |
| Trade | Duhok | 22 | 295,441,878 | 112 | 4,540,636,607 |
| | Sulaimaniya | 55 | 492,093,193 | | |
| | Erbil | 35 | 3,753,101,536 | | |
| Transportation | Duhok | 0 | 0 | 2 | 104,204,000 |
| | Sulaimaniya | 0 | 0 | | |
| | Erbil | 2 | 104,204,000 | | |
| Grand Total | Duhok | 185 | 4,951,484,944 | 713 | 41,318,711,652 |
| | Sulaimaniya | 195 | 12,519,100,250 | | |
| | Erbil | 333 | 23,848,126,458 | | |

*Sources: Analysis of data from board of investment, studies and information department, Kurdistan Region*

**Table 5. Authorized joint-ventures between foreign and Kurdistan National in Kurdistan region from 2005 to 2013- governorates analysis**

| Governorates | Total Capital Investment ($) |
|---|---|
| Duhok | 544,035,059 |
| Sulaimaniya | 2,417,699,104 |
| Erbil | 835,855,093 |
| Total | 3,797,589,256 |

*Sources: Analysis of data from board of investment, studies and information department, Kurdistan region*

**Table 6. DFI in Kurdistan region from 2005 to 2013 by country and percentage analyses**

| Country | Capital Investment ($) | Percentage |
|---|---|---|
| Egypt | 350,000,000 | 0.85 |
| Emirate | 2,527,216,000 | 6.12 |
| Georgie | 600,000 | 0.001 |
| Germany | 81,205,712 | 0.20 |
| Iran | 25,440,802 | 0.06 |
| Lebanon | 990,976,871 | 2.40 |
| New Zealand | 139,389,850 | 0.34 |
| Russia | 2,805,670 | 0.01 |
| Sweden | 13,500,000 | 0.03 |
| Switzerland | 158,665,762 | 0.38 |
| Turkey | 1,088,861,600 | 2.64 |
| Turkish | 25,000,000 | 0.06 |
| UK | 214,403,975 | 0.52 |
| USA | 127,322,925 | 0.31 |
| Iraq - Iran | 2,001,850,000 | 4.84 |
| Iraq - Canada | 2,000,000 | 0.001 |
| Iraq - Emirates | 411, 555,555 | 1. |
| Iraq - Germany | 82, 770,000 | 0.20 |
| Iraq - Jordan | 8,000,000 | 0.02 |
| Iraq - Kuwait | 51,250,000 | 0.12 |
| Iraq – Mauritius Island | 15,000,000 | 0.04 |
| Iraq – South Africa | 12,000,000 | 0.03 |
| Iraq - Sweden | 20,893,549 | 0.05 |
| Iraq - Turkey | 281,606,000 | 0.69 |
| Iraq - UK | 112,000,000 | 0.27 |
| Iraq - USA | 315,000,000 | 0.76 |
| Iraq - Netherland | 2,500,000 | 0.01 |
| Iraq – Pakistan | 13,000,000 | 0.03 |
| Iraq - Spain | 3,284,530 | 0.01 |
| Iraq- Korea - Canada | 264,670,056 | 0.64 |
| Lebanon - France | 4,160,000 | 0.01 |

**Table 7. Direct foreign investment in Kurdistan Region from 2005 to 2013**

| FDI (USD Million) | 2005 | 2006 | 2007 | 2008 | 2009 | 2010 | 2011 | 2012 | 2013 |
|---|---|---|---|---|---|---|---|---|---|
| FDI Inflow | N/A | N/A | 1,193 | 421 | 142 | 1,197 | 566 | 623 | 5,185 |

*Source: Analysis of data from board of investment, studies and information department, Kurdistan region*

Furthermore, Table 10 result indicates that within the Governorates, made up of Duhok, Sulaimaniya and Erbil, there is exist slight level of differences of inflows of FDI within the said period although the magnitude of the differences are not significant with p=0.073. Continuing Table 10 indicates that the number of inflows of Foreign Direct Investments with the period covering 2005 to 2013 across the Region is significantly different. The assertion of high level of significance in terms of number of FDIs is as a result of p = 0.0001, which is hugely below 0.05 level of significance.

**Table 8. Paired samples T-test of industrial sectors and governorates' FDI from 2005 to 2013**

| | | Paired differences | | | | | t | df | Sig. (2-tailed) |
|---|---|---|---|---|---|---|---|---|---|
| | | Mean | Std. deviation | Std. error mean | 95% confidence interval of the difference | | | | |
| | | | | | Lower | Upper | | | |
| Pair 1 | Industrial sector - amount of investment in dollars ($) | -1.059E9 | 2.058E9 | 3.296E8 | -1.727E9 | -3.922E8 | -3.214 | 38 | .003 |
| Pair 2 | Regional governorates - amount of investment in dollars ($) | -1.059E9 | 2.058E9 | 3.296E8 | -1.727E9 | -3.922E8 | -3.214 | 38 | .003 |
| Pair 3 | Industrial sector - industrial sectorial number | -11.282 | 23.188 | 3.713 | -18.799 | -3.765 | -3.038 | 38 | .004 |
| Pair 4 | Regional governorates - industrial sectorial number | -16.282 | 23.916 | 3.830 | -24.035 | -8.529 | -4.252 | 38 | .000 |

*Sources: SPSS generated Paired Samples T-test*

**Table 9. Paired samples T-test of FDI in Iraq against FDI in Kurdistan from 2005 to 2013**

| | | Paired Differences | | | | | t | df | Sig. (2-tailed) |
|---|---|---|---|---|---|---|---|---|---|
| | | Mean | Std. deviation | Std. error mean | 95% confidence interval of the difference | | | | |
| | | | | | Lower | Upper | | | |
| Pair 1 | Year - Country/Region | 2.000 | 1.859 | .537 | .819 | 3.181 | 3.728 | 11 | .003 |
| Pair 2 | Country/Region - FDI) In Million Dollars ($m) | 1.2147500E3 | 7.2244272E2 | 2.0855125E2 | -1.6737682E3 | -7.5573180E2 | - 5.825 | 11 | .000 |

*Sources: sources: SPSS generated paired samples T-test*

**Table 10. ANOVA of FDI of non-oil sectors in Kurdistan Region from 2005 to 2013**

| | | Sum of Squares | df | Mean Square | F | Sig. |
|---|---|---|---|---|---|---|
| Industrial Sector | Between Groups | 385.125 | 31 | 12.423 | .541 | .887 |
| | Within Groups | 160.875 | 7 | 22.982 | | |
| | Total | 546.000 | 38 | | | |
| Regional Governorates | Between Groups | 24.125 | 31 | .778 | 2.905 | .073 |
| | Within Groups | 1.875 | 7 | .268 | | |
| | Total | 26.000 | 38 | | | |
| Industrial Sectorial Number | Between Groups | 22005.897 | 31 | 709.868 | . | 0.0001. |
| | Within Groups | .000 | 7 | .000 | | |
| | Total | 22005.897 | 38 | | | |

*Sources: SPSS generated ANOVA*

## 4. FINDINGS

The Capital inflows from Foreign Direct Investment into Kurdistan Region of Middle East from 2005 to 2013 are huge and increasing.

The outcome of this research also shows that Capital inflows from FDI spread across the Governorates of the Kurdistan Region are significantly high.

Also found to be very high and increasing is the number of FDI across the Sectors and Governorates of the Region.

The findings show that there are significant differences in number and amount of Foreign Direct Investments within the period covering 2005 to 2013 between Iraq and Kurdistan Region of Iraq.

It is further found from the research outcomes that the level of FDI inflows into the various Sectors of the Economy in the Region within the period under consideration is not significantly different. Thus, almost all the Sectors received significant attentions, though with varying number of investments within the Governorates and Sectors.

Other relevant findings is as shown in Table 11

**Table 11. Kurdistan regional level of investment needs**

| Sector | Product | Unit of measurement | Region's estimated need | Current production level | Level of needed investment |
|---|---|---|---|---|---|
| Agriculture | Wheat | Tons | 500,000 | 500,000 | Low |
| | Oat | Tons | 600,000 | 70,000 | High |
| | Corn | Tons | 240,000 | 20,000 | High |
| | Sun Flower | Tons | 37,500 | 10,500 | High |
| | Tomato | Tons | 216,000 | 102,000 | High |
| | Cucumber | Tons | 90,400 | 68,400 | Low |
| | Eggplant | Tons | 39,600 | 20,100 | Medium |
| | Okra | Tons | 11,480 | 6,200 | Medium |
| | Onion | Tons | 56,450 | 48,900 | Low |
| | Courgette | Tons | 17,570 | 10,700 | Medium |
| | Potato | Tons | 70,000 | 61,600 | Low |
| | Grapes | Tons | 60,000 | 10,000 | High |
| | Apples | Tons | 64,000 | 11,600 | High |
| | Peaches | Tons | 25,000 | 14,100 | Medium |
| | Pomegranates | Tons | 45,000 | 15,700 | High |
| | Pears | Tons | 5,700 | 1,300 | High |
| | Apricots | Tons | 6,800 | 2,600 | High |
| | Figs | Tons | 10,900 | 3,100 | High |
| | Chicken | Tons | 98,000 | 55,000 | Medium |
| | Eggs | M | 646 | 430 | Medium |
| | Red Meat | Tons | 100,000 | 22,000 | High |
| | Milk | Tons | 400,000 | 130,000 | High |
| | Fish | Tons | 6,700 | 1,430 | High |
| | Forests | Ha | 3,000,000 | 1,280,300 | High |
| | Meadows | Ha | 1,739,000 | 1,430,000 | Low |
| Food Products and Beverage | Flour | Tons | 688,000 | 689,000 | Low |
| | Dried foods | Tons | 112,000 | 29,500 | High |
| | Potato Chips | Tons | 13,500 | 10,100 | Medium |
| | Nuts | Tons | 40,200 | 24,000 | Medium |
| | Dairy Products | Tons | 217,500 | 56,000 | High |
| | Ice Cream | Tons | 7,500 | 5,050 | Medium |
| | Salt | Tons | 80,000 | 3,700 | High |
| | Soft drinks | M Liters | 844 | 616 | Medium |
| | Mineral Water | M Liters | 496 | 537 | Low |
| | Pickles | Tons | 9,630 | 6,250 | Medium |

| Sector | Product | Unit of measurement | Region's estimated need | Current production level | Level of needed investment |
|---|---|---|---|---|---|
| | Fruit Juice | Tons | 252,000,000 | 43,000,000 | High |
| | Sesame Juice | Tons | 16,300 | 1,500 | High |
| | Spices | Tons | 22,400 | 1,900 | High |
| | Pastry | Tons | 71,300 | 34,900 | High |
| | Biscuits | Tons | 40,600 | 2,800 | High |
| | Corn Chips | Tons | 4,070 | 1,900 | High |
| | Canned Food | Tons | 20,050 | 2,700 | High |
| Paper and paper products | Tissue Paper | Tons | 10.900 | 3,900 | High |
| | Carton | Tons | 12,300 | 4,400 | High |
| | Adverting Paper | Tons | 1,640 | 900 | High |
| | Printing | Pieces | 147,500 | 58,800 | High |
| | Print newspapers and announcements | Tons | 29,500 | 20,600 | Medium |
| | Print books | Tons | 8,800 | 2,500 | High |
| | Print catalogs and calendars | Tons | 6,500 | 1,490 | High |
| Chemicals and chemical product | Filter fat | M liters | 70.6 | 48.5 | Medium |
| | Pure asphalt | Tons | 153,000 | 170,000 | Low |
| | Industrial gases | Bottles | 208,000 | 133,600 | Medium |
| | Tar | Sqm | 85,000 | 93,000 | Low |
| | Pharmaceuticals | Tons | 2,370 | 3,000 | Low |
| | Soap | Tons | 8,700 | 1,300 | High |
| | Detergents | Tons | 31,400 | 10,700 | High |
| Rubber and plastic products | Bags | Tons | 13,490 | 250 | High |
| | Doors and windows PVC | Sqm | 775,000 | 831,000 | Low |
| | Plastic tubes and pipes (PVC) | Meters | 35, 300,000 | 761,000 | High |
| | Other Plastic tubes and pipes | Tons | 56,000 | | High |
| | Baskets and plastic boxes | Tons | 8,000 | 770 | High |
| | Plastic bottles | M bottles | 452.5 | 452.6 | Low |
| | Plastic plates | sqm | 22,750 | 4,100 | High |
| | Sponge | sqm | 50,500 | 52,000 | Low |
| Non-metallic mineral products | Cut & crashed stones | Tons | 633,000 | 468,000 | Medium |
| | Sand | sqm | 676,000 | 691,200 | Low |
| | Automatic brick | M pieces | 80.5 | 99 | Low |
| | Concrete blocks | M Pieces | 181.0 | 175 | Low |
| | Curbstone | sqm | 60,000 | 73,000 | Low |
| | Ceramic and mosaic tiles | sqm | 4,540,000 | 1,935,000 | High |
| | Concrete pipes | Tons | 4,600 | 5,850 | Low |
| | Ready-mix concrete | sqm | 644,000 | 1,080,000 | Low |
| | Marble stones | sqm | 1,285,000 | 136,000 | High |
| | Ready-mix asphalt | Tons | 550,000 | 1,120,000 | Low |
| | Roof tiles | sqm | 682,000 | 110,000 | High |
| | Gypsum | Tons | 544,000 | 590,000 | Low |
| | Cement | Tons | 2,700,000 | 5,500,500 | Low |
| Basic metals | B.R.C | Tons | 10,800 | 15,000 | Low |
| | Galvanized iron | Tons | 20,550 | 11,000 | Medium |
| | Aluminum doors and windows | Tons | 31,800 | 38,000 | Low |

| Sector | Product | Unit of measurement | Region's estimated need | Current production level | Level of needed investment |
|---|---|---|---|---|---|
| | Steel reinforcements | Tons | 241,500 | 96,000 | High |
| | Iron pieces | Tons | 44,900 | 23,000 | Medium |
| | Electric Wires | Tons | 20,800 | 11,000 | Medium |
| | Steel doors and windows | Tons | 91,000 | 104,700 | Low |
| | Steel structures | Tons | 7,000 | 4,200 | Medium |
| | Water and fuel tanks | Tons | 16,100 | 16,900 | Low |
| | Iron barrels and boxes | Tons | 2,080 | 780 | High |

*Sources: KRG Board of Investment. Investment Spotlight; Issue 2 – May 2013*

## 5. CONCLUSION

The growing Tourism in Erbil and other Non-Oil Sectors has attracted and are still attracting huge Foreign Direct Investment into Kurdistan Region from 2005 to 2013. To sustain the tempo of FDI inflows into the region, Kurdistan Regional Government has to come up with a policy that will not only encourage huge local participations in banking but a Cashless economic policy. This will create more spill-over effect in attracting numerous further Foreign Direct Investment into Non-Oil Sectors of the Economy.

## 6. RECOMMENDATIONS

It is recommended that Foreign Potential Investors should concentrate on investing mainly on sectors of high level of investment needs but they require clarifications from the Kurdistan Board of Investment as to current obtainable incentives in the Sector of their interest.

Private oriented (Non-Government) Foreign Direct Investment should be started via Joint-Ventures with Kurdistan Nationals or already Kurdistan based Corporate body so as to gain adequate environmental knowledge of the business and entrepreneurial terrain of the region before split-off diversifications.

To encourage more Foreign Direct Investments into the Housing Sector, more reforms are needed in the Kurdistan Land use and acquisition law.

In order to stimulate more FDI into the Banking Sector, The Kurdistan Regional Government should enact policies and laws that will encourage and sustain a Cashless fiscal policy. This will enormously encourage local banking activities.

## ACKNOWLEDGEMENT

We acknowledge the contributions of University of Kurdistan Hewler for funding this research.

## COMPETING INTERESTS

Authors have declared that no competing interests exist.